\documentclass[amsmat,amssymb,amsfonts,aps,prb,twocolumn,showpacs]{revtex4}
\usepackage{graphicx}
\usepackage{graphics}
\usepackage{dcolumn}
\usepackage{bm}

\begin{document}

\title{Tunnel Magneto-resistance in GaMnAs: going beyond Julli\`{e}re formula.}

\author{L.Brey\footnote{brey@icmm.csic.es}}
\affiliation{Instituto de Ciencia de Materiales de Madrid (CSIC),
Cantoblanco, 28049, Madrid, Spain.}
\author{C.Tejedor}
\affiliation{Departmento de F\'{\i}sica Te\'orica de la Materia
Condensada, Universidad Aut\'onoma de Madrid, 28049 Madrid,
Spain.}
\author{J.Fern\'andez-Rossier}
\affiliation{Departamento de F\'{\i}sica Aplicada, Universidad de
Alicante, 03080 Alicante, Spain.}

\date{\today}

\begin{abstract}
The relation between tunnel magneto-resistance (TMR) and spin
polarization is explored for GaMnAs/GaAlAs/GaMnAs structures where
the carriers experience strong spin-orbit interactions. TMR is
calculated using Landauer approach. The materials are described in
the 6 band $\bf {k}\cdot \bf{p}$ model which includes spin orbit
interaction. Ferromagnetism is described in the virtual crystal
mean field approximations. Our results indicate that TMR is a
function of of spin polarization and barrier thickness. As a
result of the stong spin orbit interactions, TMR also depends on
the the angle between current flow direction and the electrode
magnetization. These results   compromise the validity of Julliere
formula.

\end{abstract}

 \maketitle

\narrowtext The relative orientation of the magnetization of two
ferromagnetic electrodes can affect dramatically electron
transport  across a tunnelling barrier connecting them. This
phenomenon gives rise to the so called tunnel-magneto-resistance
\cite{Moodera} (TMR),
${\rm TMR} =(R_{AP}-R_P)/R_{AP} $
where $R_{AP}$ and $R_P$ are  the resistances for anti-parallel and parallel
orientations respectively. TMR is exploited to fabricate devices which are
ultra-sensitive to variations of an external magnetic field
\cite{applications}. Microscopic understanding  of TMR, based upon
the hypothesis that spin is conserved in the tunnelling process, leads
to the well known Julliere  Formula\cite{Julliere,Slonczewski}:
\begin{equation}
\rm {TMR^{J}} =  \frac{2 P_L P_R}{1+P_L P_R} \label{julliere}
\end{equation}
which relates TMR with    $P_{L,R}$,  the polarizations of the
left ($L$) and right ($R$) electrodes. Assuming Eq.
(\ref{julliere}) is correct, it permits to extract the spin
polarization of the electrodes from the value of TMR in symmetric
tunnel junctions, regardless of the physical properties of the
barrier. Eq. (\ref{julliere}) can be derived doing second order
perturbation theory in tunnelling amplitude \cite{Allan}.

Diluted magnetic semiconductors like Ga$_{1-x}$Mn$_x$As are ferromagnetic below
$T_C\simeq$ 150 K  \cite{Edmonds:2002_b}. This type of materials  rise a lot of
interest because they afford the integration of ferromagnetic and
semiconducting functionalities in a single device
\cite{DMS-applic,spintronics}. Substitutional impurities of Mn in GaAs are
acceptors. The holes released by the Mn are responsible of the magnetic
ordering and transport properties of Ga$_{1-x}$Mn$_x$As. Several groups have
been able to fabricate tunnel junctions with GaMnAs in the electrodes
\cite{Tanaka,Fert,Mollenkamp} and they have reported large values of TMR
finding that its value depends on the properties  of the barrier\cite{Tanaka}.
Furthermore, spin orbit (SO) interactions for the holes in the valence band of
GaAs is very strong ($\Delta_{SO}\simeq 0.34$ eV) so that the spin of the holes
is not a good quantum number. Both the experimental results
\cite{Tanaka,Fert,Mollenkamp} and the failure of Julliere's hypothesis on
spin conservation, lead us to study TMR in GaMnAs/AlAs/GaMnAs systems using a
non-pertubative approach which fully includes spin orbit interactions. We
anticipate our main result:  in GaMnAs based heterostructures TMR  depends  on
 both the barrier thickness,   $d_b$, the angle
formed by the current flow and the magnetization.  All these features depend on
the strength of the SO coupling and are relevant for the use of GaMnAs based
heterostructures  in magnetoelectronics.

In this letter we calculate vertical transport ($z$-direction)  in
epitaxially grown GaMnAs/GaAlAs/GaMnAs tunnel systems. We take the same
Mn density at the two electrodes. The
materials are described in the 6 bands ${\bf k} \cdot {\bf p}$
model which captures the interplay of SO interaction and the
threefold orbital degeneracy  in the top of the valence bands.
Ferromagnetism is originated by the exchange interaction between
the spin of the localized Mn  and the itinerant holes.
In the case of metallic samples the mean field  and virtual
crystal (MF-VC) approximation\cite{Dietl,Ramin,jfr-sham,LB}
accounts for a number of experimental facts like the dependence of
$T_c$ on both the Mn  and hole  densities \cite{Dietl}, the
magnetic anisotropy\cite{Ramin} and magnetic circular dichroism
\cite{Dietl}. In this approach spontaneous magnetization is characterized
 by effective Zeeman magnetic field ${\cal \vec H}$ which
represents the effect of exchange interaction of the  Mn
magnetization on the spin of the holes. The MF-VC approximation
restores the translational invariance in the Hamiltonian, and it
is possible to label the eigenvectors and eigenvalues by a band
index ,$n$, and a wavevector $(k_z,{\bf k}_\parallel)$. Due to the
SO interaction, energy bands and Bloch states depend on both the
magnitude and the orientation of ${\cal \vec H}$, and spin is {\em
not} conserved. For a given set of parameters representing GaMnAs,
a  magnetic orientation $\vec{\Omega}$ and a temperature, the
above approach yields the carrier spin polarization $P$. In this
model, the electronic structure of GaMnAs is fully characterized
by $P$, the hole density $p$ and $\vec{\Omega}$. For a fixed
Mn-hole exchange coupling strength, $p$ and $\vec{\Omega}$ there
is a one to one correspondence between the Mn concentration and
$P$. We describe the valence bands of the barrier (GaAlAs) with
the same parameters used for GaAs, but
offset\cite{Wessel,Chuang,Lius} a potential $V_b$ of the order of
300meV.

The conductance is calculated using the Landauer formula
\cite{Wessel,Chuang,Lius},
\begin{equation}
G_{\vec{\Omega}_L,\vec{\Omega}_R}=\frac{e^2}{2 \pi \hbar}\sum_{n,n',{\bf
k}_\parallel} T _{n,n'} ^{{\bf k} _\parallel} (E_F) \label{G}
\end{equation}
where $E_F$ is the Fermi energy,  $T _{n,n´} ^{{\bf k} _\parallel}
(E_F)$ is the transmission probability from a state in the left with band index
$n$ to a state in the right with band index $n'$, and
$\vec{\Omega}_{L,R}$ are the magnetization orientation of the electrodes.
Only the energy and
${\bf k} _\parallel$ are conserved in the tunnelling process. The transmission
matrix  is obtained in the transfer matrix method applied to the 6 bands ${\bf
k} \cdot {\bf p}$ hamiltonian\cite{Wessel,Chuang,Lius} including the exchange
field ${\cal \vec H}$ \cite{Petukhov}.

\begin{figure}
\includegraphics[clip,width=8cm]{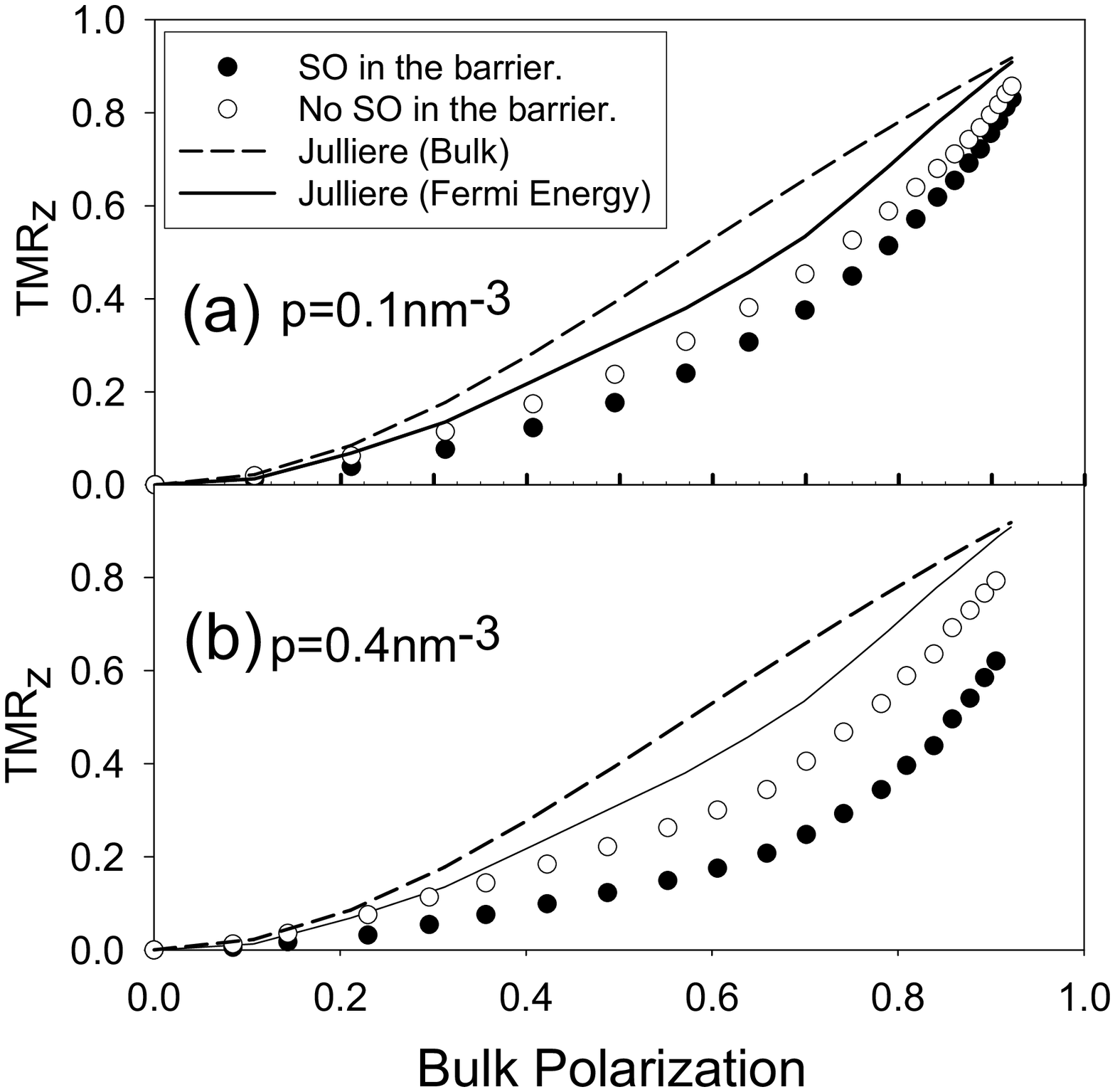}
\caption{TMR$_z$ as function of $P$.  $d_b$=15$\AA$ and
$V_b$=300meV.  Case (a)p=0.1nm$^{-3}$ and  (b)p=0.4nm$^{-3}$. }
\label{fig1}
\end{figure}

Conventionally, $G$ depends on the angle formed by $\vec{\Omega}_L$ and
$\vec{\Omega}_R$. In presence of SO coupling the conductance depends also on the
angle formed by the current with the magnetization. This makes it necessary to
define  both TMR$_z$,  for the case of current parallel (or antiparallel)  with
the magnetization, and TMR$_x$,  for the case where the magnetizations are
perpendicular to the current flow:
\begin{equation}
\rm{TMR_z} = \frac{G_{\uparrow,\uparrow} -
G_{\uparrow,\downarrow}}{G_{\uparrow,\uparrow}} \, \, \, \, \, \,
\, \, \, \rm{TMR_x} = \frac{G_{\leftarrow,\leftarrow} -
G_{\leftarrow,\rightarrow}}{G_{\leftarrow,\leftarrow}}\, \, .
\nonumber
 \label{G_direc}
\end{equation}
where we denote the positive (negative) $z$-direction as $\uparrow
(\downarrow)$, and the positive (negative) $x$-direction as
$\leftarrow (\rightarrow)$.

In Fig.1 we plot  TMR$_z$  for a symmetric GaMnAs/AlAs/GaMnAs
junction as function of the bulk spin polarization $P$ for two
hole densities $p$=0.1nm$^{-3}$ and $p$=0.4nm$^{-3}$.  The barrier
height is $V_b=$300 $meV$. Both the Julliere expression and
TMR$_z$   are  increasing functions of the electrode spin
polarization $P$. However for intermediate values of the
polarization and $p$=0.4nm$^{-3}$ the difference between the
Julliere and the calculated TMR can be of a factor of two. The
discrepancy remains if we use the Fermi surface polarization
\cite{Mazin} $P^{FS}$ instead of the bulk polarization $P$ . In
order to address whether the discrepancy is due to the SO
interaction, we also show in figure 1 TMR$_z$ in the case where
the SO interaction is suppressed in the barrier. Interestingly,
$TMR_z$ is larger when SO is suppressed and closer to  the
Julliere result, but still off.
\begin{figure}
\includegraphics[clip,width=8cm]{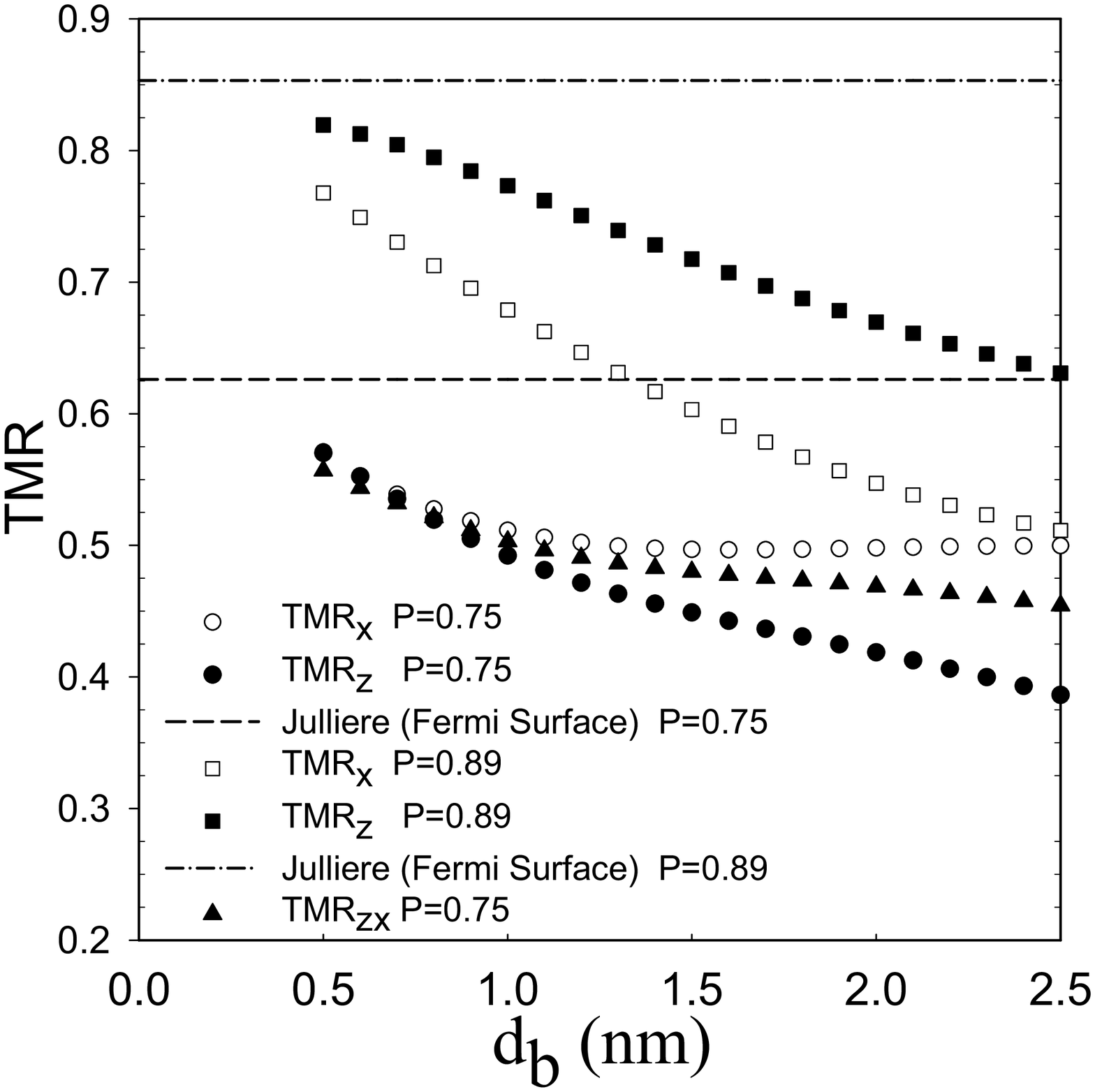}
\caption{TMR$_z$ and TMR$_x$ as function of $d_b$ for $P$=0.75 and
$P$=0.89. $V_b$=300meV and $p$=0.1nm$^{-3}$.} \label{fig2}
\end{figure}

In Fig.2 we plot both TMR$_z$ and TMR$_x$ as a function of the
barrier thickness $d_b$ for two different values of $P$.
 The Julliere value, independent of $d_b$ is also shown.
Due to the strong SO
coupling in the system, TMR$_z$ and TMR$_x$ can be quite
different, and because of the complicated matching at the
interface, its relative magnitude depends strongly on the
parameters of the system. In agreement with
experiments\cite{Tanaka}, and tight-binding
calculations\cite{Moodera,Tanaka} we find that TMR decreases
rapidly for thin barriers, in marked contrast with Eq. (\ref{julliere}).
We also plot the TMR when the magnetization is oriented
in the [101] direction, TMR$_{zx}$. This quantity turns out to be
almost exactly the  average of TMR$_{x}$ and TMR$_{z}$, indicating that the
experimental\cite{Tanaka} difference in the TMR between the [100]
and [110] field directions is due to the SO coupling and not to
the cubic magnetocrystalline anisotropy in the GaAs crystal
structure.

A quantitave comparison between our results and
the experiments is difficult, because
 quantities like the carrier density and band offset are
unknown in the heterostructures. Even
$d_b$ is not accurately known \cite{Tanaka}. Qualitatively,
though,  our results  account for
the decreases of TMR with the barrier thickness \cite{Tanaka}, and also show a
strong dependence of the TMR values on the angle formed by the
magnetization and the current flow.

Impurity scattering combined with the strong SO coupling produces
anisotropy in the DC transport properties of {\em bulk}
GaMnAs\cite{amr}. We have analyzed the variation of the ballistic
tunnelling resistance when the current flow is parallel or
perpendicular to the magnetization. In Fig.3  we plot the
anisotropy tunnelling magneto resistance (ATMR),
\begin{equation}
\rm{ATMR} = \frac{G_{\uparrow,\uparrow} -
G_{\rightarrow,\rightarrow}}{G_{\uparrow,\uparrow}} \,\, \, .
\end{equation}
The tunnelling current can change up to 6 $\%$ for large values of
$P$. This number is comparable with those obtained for bulk
\cite{amr}. Therefore, in such a heterostructure, the ATMR can add
a significant contribution to the bulk anisotropy magneto
resistance of the electrodes.

\begin{figure}
\includegraphics[clip,width=8cm]{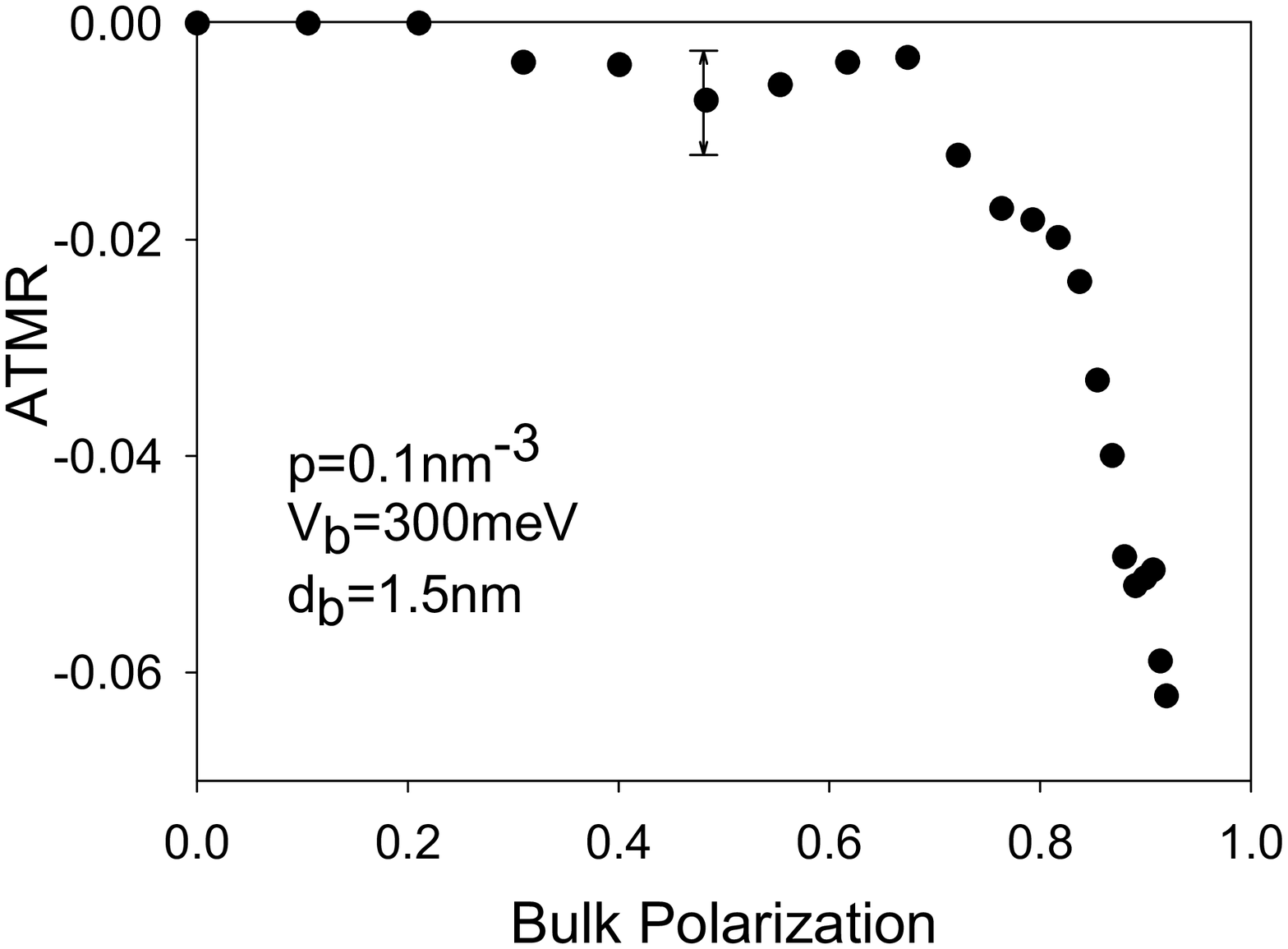}
\caption{Anisotropic TMR as function of the bulk polarization. The
error bar is an estimation of the numerical error in the calculations.}
\label{fig3}
\end{figure}

In conclusion, the presence of strong spin-orbit interaction modifies the
conventional relation between TMR and spin polarization of the electrodes.  As
in the conventional case,  our results show that  TMR is an increasing function
of the electrodes polarization $P$, quantitatively different  than Eq.
(\ref{julliere}). Large values of TMR  can be reached for sufficiently high $P$
even in the presence of strong spin orbit interaction.  In contrast with Eq.
(\ref{julliere}), TMR depends on the barrier thickness.   Therefore,
 (\ref{julliere}) is
not an appropriated tool  for inferring the spin polarization of the electrodes
in GaMnAs based heterostrucures. Finally, because of the spin orbit interaction,
 both the TMR and the conductance depend on the angle between the current and
the magnetization orientation. These are  qualitatively new  physical phenomena
that might be exploited to build spin valves with new functionalities.


Work supported in part by MCYT of Spain under contract Numbers
MAT2002-04429-C03-01, MAT2002-00139, MAT2003-08109-C02-01,
Fundaci\'on Ram\'on Areces, Ramon y Cajal program and UE within
the Research Training Network COLLECT.

\newpage
{\underline{Figure Captions}}

\vspace{1.cm}

 \noindent Figure 1: TMR$_z$ as function of $P$.
$d_b$=15$\AA$ and $V_b$=300meV.  Case (a)p=0.1nm$^{-3}$ and
(b)p=0.4nm$^{-3}$.

\vspace{1.cm}

\noindent Figure 2: TMR$_z$ and TMR$_x$ as function of $d_b$ for
$P$=0.75 and $P$=0.89. $V_b$=300meV and $p$=0.1nm$^{-3}$.

\vspace{1.cm}

\noindent Figure 3: Anisotropic TMR as function of the bulk
polarization. The error bar is an estimation of the numerical
error in the calculations.

%
%
%
%
%
%
%
%
\end{document}